# Transient Analysis during Maximum Power Point Tracking (TrAMPPT) to Assess Dynamic Response of Perovskite Solar Cells

Aniela Czudek[1a,2], Katrin Hirselandt[1a], Lukas Kegelmann[1b], Amran Al-Ashouri[1b], Marko Jošt[1b], Weiwei Zuo[1c], Antonio Abate[1c], Lars Korte[1d], Steve Albrecht[1b], Janardan Dagar[1a], Eva L. Unger[1a,3]

[1] Helmholtz-Zentrum Berlin, HySPRINT Innovation Lab, Kekuléstrasse 5, 12489 Berlin, Germany

[a] Young Investigator Group Hybrid Materials Formation and Scaling

[b] Young Investigator Group Perovskite Tandem Photovoltaics

[c] Young Investigator Group Active Materials and Interfaces for Stable Perovskite Solar Cells

[d] Institute for Silicon Photovoltaics

[2] Faculty of Physics, Warsaw University of Technology, Koszykowa 75, 00-662 Warsaw, Poland

[3] Department of Chemistry & NanoLund, Lund University, Naturvetarvägen 14, 22362 Lund, Sweden

## Abstract

Determination of the device performance parameters of perovskite solar cells is far from trivial as transient effects may cause large discrepancies in current-voltage measurements as a function of scan rate and pre-conditioning. Maximum power point tracking, *MPPT*, enables to determine the steady-state maximum power conversion efficiency. However, the *MPPT* does not provide any information on the device performance parameters, which are reliable only if extracted from current-voltage curves collected under steady-state conditions. We show that is possible to determine the shorter settling or delay time suitable to carry out *J-V* measurements under steady-state conditions by analysis of the transient device response around the MPP. This procedure proves to be more time-efficient than measurement *J-V* measurements at a variety of scan rates. Furthermore, the generic algorithm presented here can be implemented to assess changes in the dynamic response of devices during long-term device ageing

# 1. Introduction

Recent progress in solar cells based on metal-halide perovskites has demonstrated that the know-how in various different types of solar cell devices can lead to rapid progress upon discovery of novel semiconductors suitable for solar energy conversion.[1,2] The development of solar cell technology requires reliable device characterization tools which yield performance data that is representative of the steady state device operation. Current density - voltage, $J$-$V$, measurements are the most commonly used method to assess the device performance of solar cells. Debates regarding hysteresis and slow transient phenomena in the $J$-$V$ response of perovskite solar cells have highlighted that the power conversion efficiency and maximum power point, $MPP$, derived from these measurements might not be representative for the steady-state performance of devices.[3–5] Transient electronic phenomena in perovskite devices have been connected to the re-distribution of ionic charge carriers.[3,6–11] These are interpreted to cause changes in the internal electric field distribution affecting the charge carrier extraction efficiency and recombination rates rendering the photocurrent dependent on scan rate and direction.[12,13] Apart from changes in the electric field distribution, charge carrier trapping/de-trapping effects within interfacial trap states[14–18] could contribute to transient capacitive phenomena. Uncertainties in determining the $MPP$ from $J$-$V$ measurements has prompted the recommendation for reference measurements to account for the transient device response,[4,19,20] and some measures to verify device performance metrics are now often also requested by scientific journals.[21] These often also include the recommendation to provide measurements more representative of the steady-state response of perovskite-based solar cells by for instance monitoring the $MPP$ over time. Device performance metrics of short circuit current, $J_{SC}$, open circuit voltage, $V_{OC}$, and fill factor, $FF$, should be derived from measurements representative of steady-state conditions, which is often difficult to define for perovskite solar cells.

Numerous reports have proposed to quantify the discrepancy between the forward and reverse $J$-$V$ response and various equations calculating a measure for the difference between these two $J$-$V$ curves, expressed as hysteresis indices, $HIs$.[16,22–24] As an example, equation 1 defines a commonly used hysteresis index taken as the difference between the integrated reverse (open-circuit to short-circuit) and forward (short-circuit to open-circuit) $J$-$V$ curve, normalized to integral over the reverse-scan $J$-$V$ curve, between 0 V and $V_{OC}$.

$$HI = \frac{\int_{V_{OC}}^{0\,V} J_R(V) - \int_{0\,V}^{V_{OC}} J_F(V)}{\int_{V_{OC}}^{0\,V} J_R(V)} \qquad (1)$$

Hysteresis indices are, however, not sufficient criteria by themselves, as the presence or absence of hysteresis – discrepancy between *J-V* scans in different scan directions – is strongly dependent on the scan rate. If a *J-V* measurement is performed at a much higher scan rate than the device response time,[25] the curves measured in forward and reverse direction may exhibit good overlap[3] and hence a low *HI*. Hysteresis indices indeed are of limited scientific significance as these metrics might be representative of the device response in non-equilibrium conditions.[3,25,26] However, performing *J-V* measurements by default at very slow scan rates in order to reach a quasi-equilibrium is also not a viable solution to ensure reliable measurements as irreversible degradation may overshadow the benefits of reducing transient influence, as recently addressed by Dunbar et al.[27] Furthermore, introducing increasingly complex and time-consuming measurement protocols is counter-productive as device degradation may obscure their performance characterization.

What is needed are analytical tools probing different time scales of transient phenomena in perovskite devices that need to be developed to gain insight into and disentangle transient phenomena of different origins. Ideally, measurements should capture the transient response of a device on different time scales to enable the distinction between reversible capacitive phenomena and irreversible changes in the device due to degradation.[28,29] Device performance should be assessed using maximum power point tracking,[30,31] *MPPT*, which could and maybe should become the more relevant metric when assessing perovskite solar cells performance rather than the power conversion efficiency, *PCE*, or maximum power point, *MPP*, determined from current density-voltage, *J-V*, measurements. Keeping the device around the maximum power point, the influence of transient effects during *J-V* cycling is omitted and changes in device performance due to pre-conditioning can be accounted for while monitoring device performance over a longer time period.

There are different methods to carry out maximum power point measurements over time. In some cases, the current density at a fixed voltage determined from the *MPP* of *J-V* measurements, $V_{mpp}$.[28,32–34] Preferably, the time-evolution of the *MPP* should be assessed by perturb-and-observe *MPPT* algorithms that implement a periodic perturbation of the applied voltage maximizing the total power output of the device as $V_{mpp}$ may change over time.[30,31,35–

[37] For perovskite solar cells, perturb-and-observe *MPPT* algorithms that dynamically adjust the sampling time and voltage step size have been proposed as the dynamic response of devices changes due to changes in internal electric field distribution and capacitance during device operation.[30,31,36] Cimaroli et al.[30] proposed a *predictive MPPT* algorithm that derives the steady-state power by fitting the current response to a voltage perturbation with a biexpontential function. However, even perturb-and-observe as well as *predictive MPPT* algorithms might get stuck in local performance maxima when devices exhibit current-voltage hysteresis.[31,36] Pellet et al.[31] therefore proposed a so-called *hill-climbing MPPT* algorithm starting at applied potentials larger than the *MPP*, which was found to have a positive effect on device performance.

We here present a novel approach that to some extent reverses the order of measurements usually performed to assess the *MPP* of solar cell devices: starting with an *MPPT* measurement and ending with *J-V* measurements at scan conditions determined from the transient device response assessed during *MPPT*. We introduced a deliberate voltage perturbation around *MPP* as part of a perturb-and-observe *MPPT* algorithm to record the transient current density response of the device, as illustrated in Scheme 1. The transient current response was analyzed by bi-exponential fits according to transient current response analysis discussed elsewhere,[38,39]

$$J(t) = J_{SS} + A_{fast}e^{-\left(\frac{t-t_0}{\tau_{fast}}\right)} + A_{slow}e^{-\left(\frac{t-t_0}{\tau_{slow}}\right)} \qquad (2)$$

from which the transient time constants, $\tau_{fast}$ and $\tau_{slow}$, amplitudes $A_{fast}$ and $A_{slow}$, as well as the steady state current, $J_{SS}$, can be determined. We show that the slow transient time constant, $\tau_{slow}$, is a suitable minimum value to be used as delay or voltage settling time to perform a single current density-voltage, *J-V*, measurement that exhibits minimum hysteresis between forward and reverse scan direction. As our procedure captures the MPP as a function of time as well as dynamic response of devices under investigation, we refer to it as "transient analysis during maximum power point tracking", *TrAMPPT*.

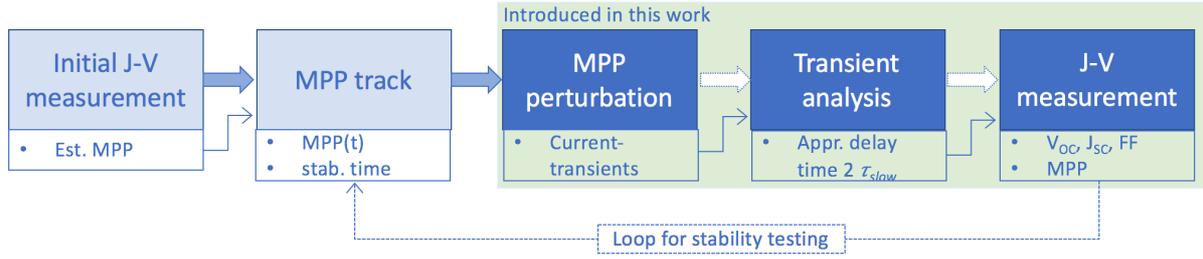

**Scheme 1:** Flowchart of proposed Transient Analysis during Maximum Power Point Tracking (*TrAMPPT*) algorithm to determine transient solar cell response from voltage perturbation during MPPT measurements. The first two steps present a common MPPT algorithm of a perturb-and-observe MPP tracking measurement following an initial *J-V* measurement. Presented here is the inclusion of a deliberate MPP perturbation to assess the transient response of the device under investigation. From the analysis of the transient current response, a minimum delay time, $t_{delay}$, can be determined from the slow transient time constant $\tau_{slow}$. This can be used as a suitable input parameter for *J-V* measurements under steady-state conditions.

We find that this measurement strategy offers an elegant and time-efficient solution to assess the MPP as well as determine measurement conditions to perform *J-V* measurement under quasi-steady-state conditions of the device that allows the determination of the device performance metrics of short-circuit current, $J_{SC}$, open circuit voltage, $V_{OC}$, and fill factor, *FF*, unperturbed by transient effects. These metrics are important to assess the performance and identify losses of devices relative to their thermodynamic limit, which is not directly be assessed by *MPP* alone.

The generic algorithm could be implemented as long-term MPPT assessment of devices using the periodic determination and analysis of transients to capture possible changes in the dynamic device response due to aging effects. This enable the determination of suitable delay times, $t_{delay}$, to perform *J-V* measurements appropriate for the device under investigation at the specific point in time during its lifetime.

## 2. Experimental

### 2.1 Dynamic Maximum Power Point Tracking Algorithm

The maximum power point tracking algorithm was based on a standard perturb-and-observe *MPPT*[40,41] measurement routine implemented as part of our current density – voltage, *J-V*, LabVIEW based measurement program in our laboratory.[42] To avoid getting stuck in local performance minima, an estimate for the *MPP* is derived from a quick initial *J-V* measurement. The regular algorithm perturbs the applied voltage by a double step of +/- 10 mV around the maximum power point voltage, $V_{MPP}$, compares the solar cell's output power at these three voltages and then sets the new $V_{MPP}$ to the one corresponding to maximum power. It is important that the step duration is set long enough for transients to equilibrate before the power is calculated at the newly set voltage level.[31]

We expanded this standard procedure by introducing a voltage perturbation phase to monitor the transient current density response of devices, indicated in Scheme 1. As the amplitude of the current transient is a function of the voltage step size, we usually chose a +/- 50 mV double step around $V_{MPP}$ for the voltage perturbation, leading to a large enough amplitude for the fitting procedure. The transient current response during *MPP* perturbation for the voltage steps to $V_{MPP}$ was analyzed by fitting with equation (2). This allows the extraction and comparison of characteristic fast and slow transient time constants, $\tau_{fast}$ and $\tau_{slow}$, amplitude of transient current response, $A_{fast}$ and $A_{slow}$, as well as the steady-state current, $J_{SS}$.

The slow transient time constants, $\tau_{slow}$, give a measure for appropriate delay times, $t_{delay}$, that are suitable to perform current density – voltage, *J-V*, measurements at steady-state conditions. This is equivalent to waiting "long enough" after a voltage step to let the transient current response decay towards steady-state. As indicated in Scheme 1, in this work the acquisition and analysis as well as consecutive *J-V* measurements were carried out as separate steps.

**2.2 Devices investigated**

Measurements presented herein were carried out on p-i-n and n-i-p devices prepared in baseline manufacturing of metal-halide perovskite solar cells in the HySPRINT laboratory. We here compare p-i-n and n-i-p thin film architecture types, where p and n stand for p- and n-type selective contact layers and i for the perovskite layer, assuming that it can be considered an intrinsic semiconductor. For both devices $Cs_{0.05}(FA_{0.83}MA_{0.17})_{0.95}Pb(I_{0.83}Br_{0.17})_3$ perovskite, sometimes referred to as "triple cation" perovskite reported by Saliba et al.[43], was utilized as the light-harvesting layer and deposited by spin-coating from a precursor solution as described in more detail in the supporting information. For the p-i-n device, poly[bis(4-phenyl)-(2,4,6-

trimethylphenyl)amine (PTAA) was used as the p-type selective contact layer on transparent conducting indium-doped tin oxide (ITO) on glass substrates. Consecutively evaporated layer of $C_{60}$, bathocuproine (BCP) and silver were used as n-type selective contacts. For the n-i-p device, spin-cast tin oxide ($SnO_2$) on ITO served as n-type selective contact and Li-bis(trifluoromethanesulfonyl) imide (Li-TFSI) doped 2,2',7,7'-tetrakis-(N,N-di-p-methoxyphenylamine)-9,9'-spirobifluorene (spiro-OMeTAD) contacted with gold (Au) as p-type selective contact layers. Detailed information all processing steps can be found in the Supporting Information.

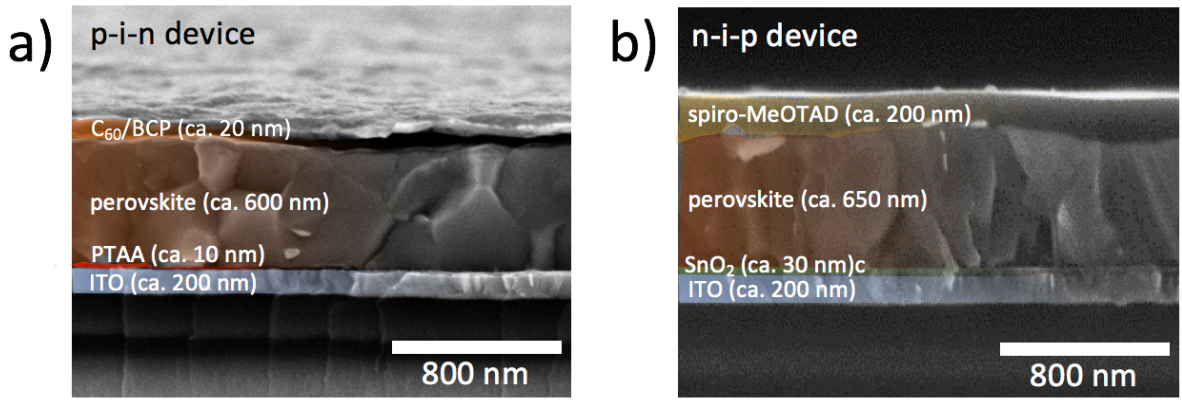

**Figure 1:** SEM-cross section images with faux color overlays (drawn manually in power point) to illustrate device architectures of the a) "inverted" p-i-n device and b) "regular" n-i-p device investigated in this work. In both cases, the "triple cation" $Cs_{0.05}(FA_{0.83}MA_{0.17})_{0.95}Pb(I_{0.83}Br_{0.17})_3$ perovskite was used as absorber. The p-i-n device comprises PTAA as a p-type selective contact on conductive ITO and $C_{60}$/BCP/Ag as n-type back contact. The n-i-p device comprises $SnO_2$ as n-type selective contact and spiro-MeOTAD/Au as p-type selective contact.

## 2.3 Current density-voltage measurements

*TrAMPPT* and *J-V* measurements were performed either in a nitrogen-filled glovebox (p-i-n devices) or in ambient atmosphere (n-i-p devices). Measurement on p-i-n device (inside the glovebox) were carried out using an Oriel LCS-100 class ABB solar simulator while measurements outside were carried out using a Wavelabs Sinus-70 Class AAA sun simulator. Both were calibrated to AM1.5G standard using a calibrated silicon reference diode (Fraunhofer ISE). A 3.4 mm x 3.4 mm shadow mask was used to measure on individual pixels defining our active area to 0.1156 $cm^2$. A quick initial *J-V* measurement was performed on all pixels, scanning from -0.2V to 1.2V and back from 1.2V to -0.2V with 50 mV voltage steps, delay time of 20 ms and integration time of 40 ms. The *TrAMPPT* measurement and analysis was carried out on the best performing pixel within the data set while other pixels were masked to avoid crosstalk. Here, the *TrAMPPT* measurements were performed for a total time span of 500 s but the duration of the MPP tracking and MPP perturbation phase can be adjusted by the user. To compare the transient response determined from *TrAMPPT* measurements, we also carried out *J-V* measurements at different scan rates, as specified by different delay times as shown in Table S1. As a standard, we carried out forward scans (F), from $V \leq 0V$ towards $V \geq V_{OC}$, followed by reverse scans (R) in the opposite direction. The *J-V* discrepancy was analyzed using the definition of the hysteresis index, *HI*, according to equation (1).

## 3. Results and Discussion

We will here present and compare *TrAMPPT* measurement results for two different device types: a p-i-n and an n-i-p device to showcase that this proposed measurement procedure can be employed to both. Details on the device architecture and layer stack can be found in section 2.2 and the supporting information. Section 3.2 is dedicated to the analysis of the difference in transient device response according to equation (2) while section 3.3 discusses *J-V* measurements carried out at delay times, $t_{delay}$, determined from transient analysis. Section 3.4 compares information on the transient device response gathered during *TrAMPPT* measurements in comparison with *J-V* discrepancy expressed as hysteresis indices, *HI*, according to equation (1) as a function of delay time.

### 3.1 Maximum Power Point Tracking (MPPT) Measurements

As outlined in section 2.1 and 2.3 as well as shown schematically in Scheme 1, the *TrAMPPT* measurement scheme initiates with a quick initial *J-V* scan to determine a starting value for $V_{MPP}$. In our measurements, we can usually discern four distinct phases. The device characteristics during *MPP* tracking are marked as phases I-III in Figures 2a and 2b, while the phase marked IV is the results of the deliberate voltage perturbation with 50 mV steps as apparent from the panel showing the applied voltage. We distinguished phase I and II from the more "steady" device response during *MPP* tracking as I sometimes exhibits a dramatic change in the photocurrent reflecting capacitive effects as the device is held at open circuit conditions immediately prior to stepping to $V_{MPP}$ for the *MPP* tracking measurement. Phase II marks an initial phase in which the $V_{MPP}$ equilibrates from a value determined from a fast *J-V* measurement to a value more representative of the device under steady state conditions. This change of $V_{MPP}$ over time underlines the need to utilize a perturb and observe algorithm and measurements at a fixed $V_{MPP}$ derived from fast *J-V* measurements are not representative of the *MPP* of the device under operation conditions. Phase III marks a fairly steady state device response, which for the p-i-n device is considerably more monotonous, considering the $V_{MPP}$, compared to the n-i-p device.

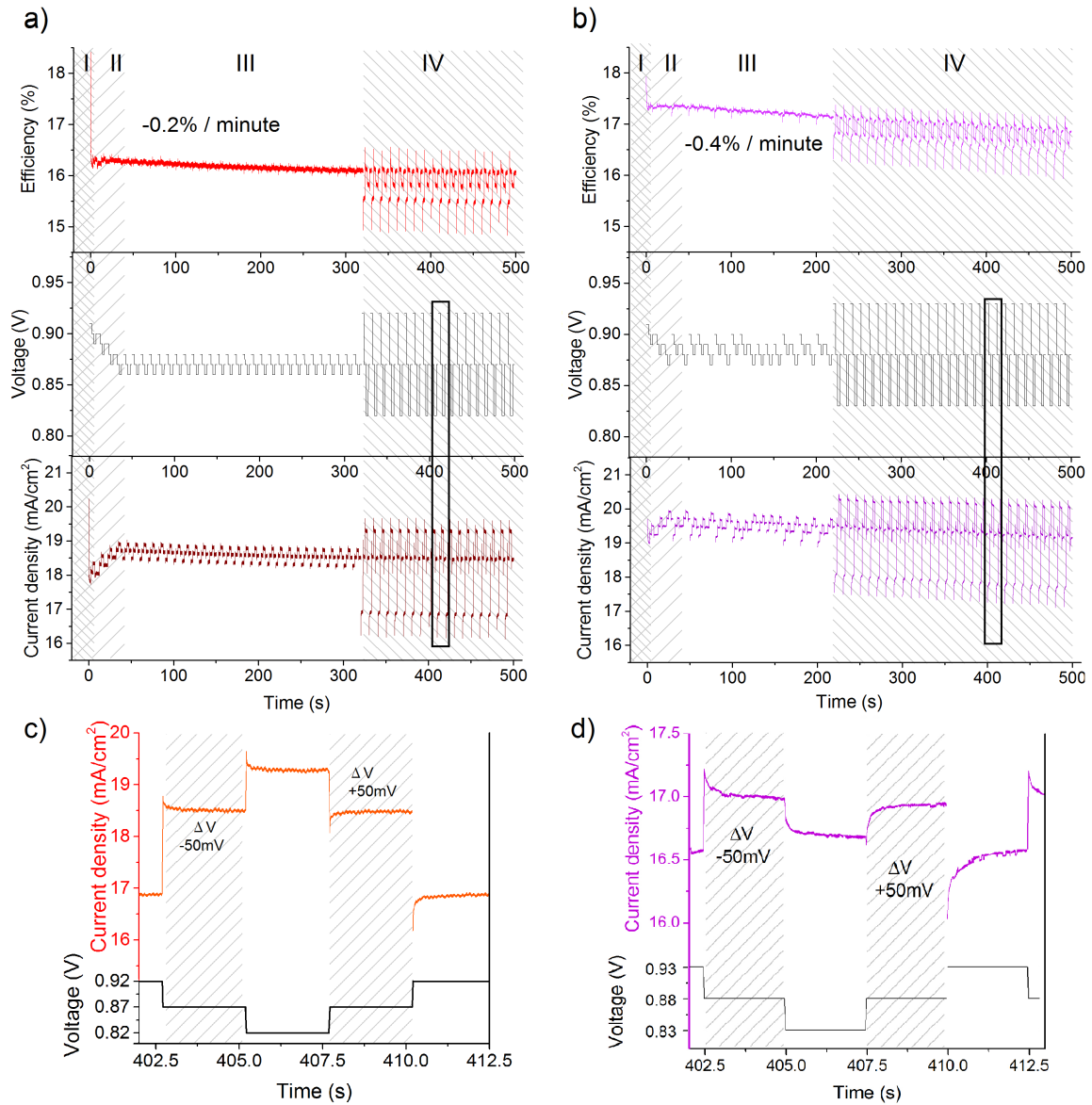

**Figure 2:** *DynaMPPT* **measurement of a p-i-n (a) and a n-i-p (b) device in comparison. The top panels show the temporal evolution of power conversion efficiency (equivalent to maximum power point, MPP), maximum power point voltage $V_{MPP}$ and current density, $J_{MPP}$. Four regions are marked with I) exhibiting current drop due to stepping to $V_{MPP}$, II) an equilibration phase in which the $V_{MPP}$ adjusts and III) a phase where the MPP appears to be quite stable. In phase IV, a deliberate voltage perturbation around $V_{MPP}$ by a double step of +/- 50 mV was introduced to investigate the dynamic response of the device current to this**

perturbation. (c) and (d) show a close-up of one perturbation cycle – black boxes in (a), (b) - for p-i-n (c) and n-i-p device (d).

During the MPP tracking phase (phase III) the device efficiency was determined to be 16.3 % with the current density declining from 18.7 mAcm$^{-2}$ to 18.5 mAcm$^{-2}$ indicating an average degradation of 0.2 mAcm$^{-2}$/270 s, i.e. a relative decrease in performance of *ca.* -0.2% per minute. The n-i-p device exhibited a MPP performance of 17.3 with a relative performance decrease of *ca.* -0.4% per minute.

During the 50 mV voltage perturbation phase (phase IV) of the *TrAMPPT* measurement, the transient response of the photocurrent becomes very pronounced. A close-up of the data is shown in Figure 2 c) and d). The current responds with an over- or under-shoot followed by an exponential decay upon a change in the applied potential. Already from the enlarged section of Figure 2 (c and d), it is apparent that the transient response appears to be slightly slower for the n-i-p compared to the p-i-n device, which will be discussed in greater detail in the next section. Transient time constants are found to be comparable for voltage perturbations of different magnitude, however the amplitude changes as a function of voltage step size as commented in Figure S5.

The p-i-n device investigated here has a lower performance compared to current state-of-the-art baseline devices with higher PCEs, achieved by further optimization of selective contact layers.[44,45] These devices show negligible hysteresis for all scan rates. The p-i-n device discussed here was a device that was aged for three days and chosen as an example for a device of this architecture type that does exhibit hysteresis, particularly for faster scan rates, as will be further discussed in section 3.3. The n-i-p device presented here represents an improvement with respect to previous results obtained for a similar device architecture.[46]

### 3.2 Transient Current Analysis upon Voltage Perturbation

Figure 3 (a) and (b) shows the transient photocurrent response for all voltage perturbation cycles towards $V_{MPP}$, which are indicated as patterned shading in Figure 1 (c) and (d). We define the current density transients for the voltage steps from -50 mV to $V_{MPP}$ as $J_{trans,F}(t)$, as the scan direction is equivalent to a forward voltage sweep, and transients observed upon changing the voltage from +50 mV to $V_{MPP}$ as $J_{trans,R}(t)$, as this coincides with a reverse voltage scan direction.

Right after the voltage step at early times, the photocurrents differ strongly for steps upwards or downwards in voltage. For the devices investigated, the photocurrent transients are found to be symmetric for the - 50 mV and + 50 mV steps towards $V_{MPP}$ indicating that the device is affected in a similar manner by the +/- voltage perturbation. Current transients merge for longer settling times indicating that a similar steady state current density, $J_{SS}$, is reached.

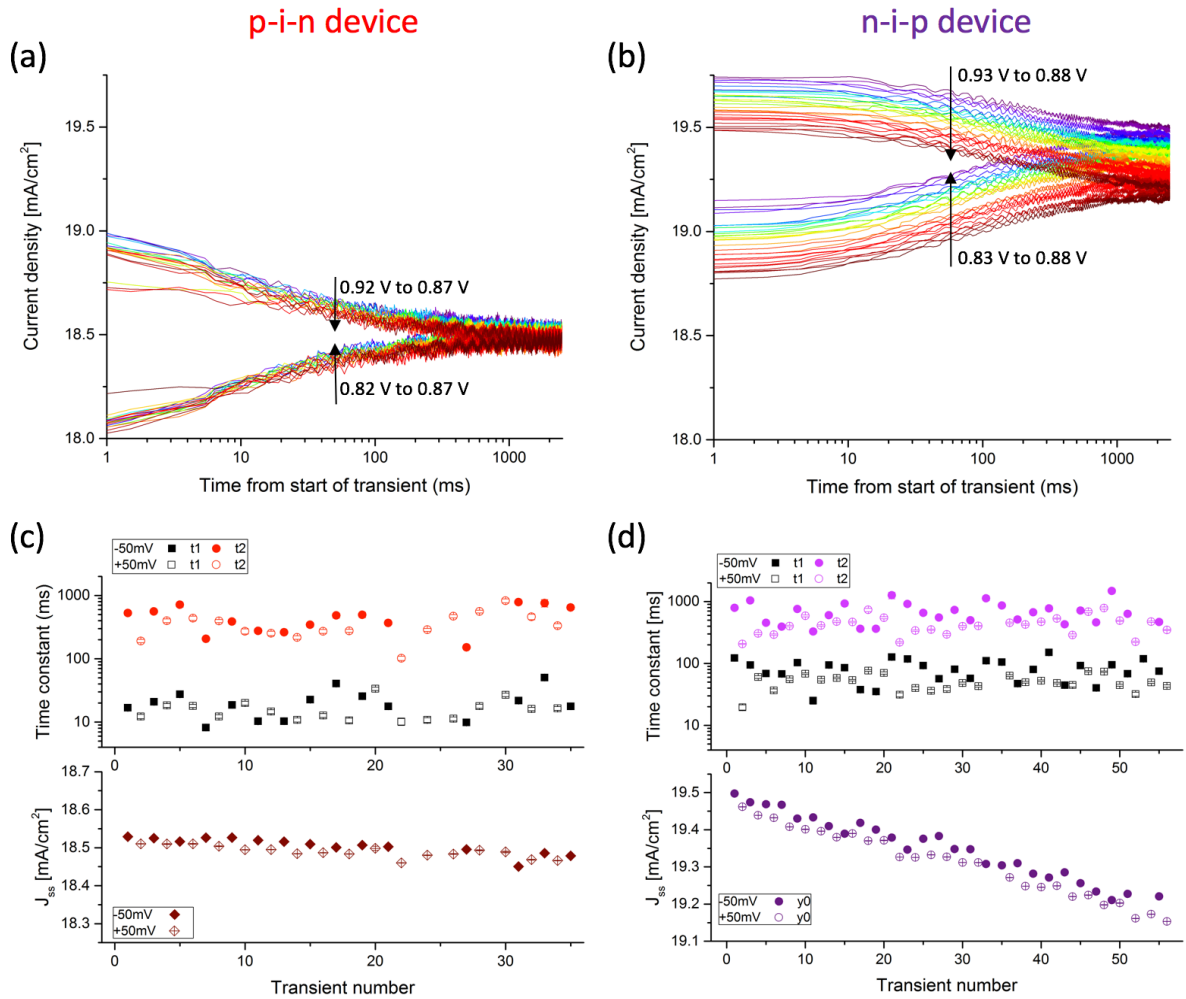

**Figure 3: Transients upon positive and negative voltage steps (+ 50mV and - 50mV respectively) during MPP perturbation for the p-i-n (a) and n-i-p (b) devices. Some oscillations of 10 Hz are observed and due to noise from the power source. Transient time constants $\tau_{fast}$ and $\tau_{slow}$ as well as steady-state current, $J_{SS}$ determined from bi-exponential fits of transients according to equation (2) for the p-i-n (c) and n-i-p (d) device.**

The transient response for both devices were fitted with equation (2) and the extracted time constants, $\tau_{fast}$ and $\tau_{slow}$, and $J_{SS}$ are shown in Figure 3 (c) and (d). The $J_{SS}$ for the p-i-n device remains fairly similar for all perturbation cycles while for the n-i-p device, a distinct decline in $J_{SS}$ can be observed.

The fact that more than one transient time constant is needed to fit the devices transient behavior, may indicate that there is more than one underlying cause for the transient response and hence for hysteresis in perovskite solar cells. For both device polarities investigated here, the transients appear to be mirror images of each other, suggesting that the underlying cause can be considered due to capacitive charging/discharging phenomena.

For the p-i-n device, the time constants and amplitudes are identical opposites within experimental error. Figure 3 c) shows the extracted $\tau_{fast}$ and $\tau_{slow}$ from fits from equation (2) to all transients showing that the +/- 50 mV step exhibit similar values. Average time constants of 18 (±9) ms for $\tau_{fast}$, and 410 (±190) ms for $\tau_{slow}$ were determined and the steady-state current $J_{SS}$ was found to decrease from 18.52 mA/cm$^2$ with 0.019 mA/cm$^2$min. The n-i-p device (Figure 3 d) exhibits average time constants of 63 (±25) ms for $\tau_{fast}$, and 540 (±230) ms for $\tau_{slow}$ and a $J_{SS}$ of 19.48 mA/cm$^2$ decreasing by 0.065 mA/cm$^2$min. The two devices investigated exhibit a comparable slow component of the transient response but the fast component is markedly slower for the n-i-p device compared to the p-i-n device.

For both devices, we observe an initial difference of about 1 mA in the current density around MPP, $J_{MPP}$, between the + 50 mV and – 50 mV step this amounts to a discrepancy in absolute performance of about 5% when measurements would be carried out at very short $t_{delay}$. This will be further discussed in section 3.4.

## 3.3    Current density – voltage, *J-V*, measurements

From the transient analysis discussed in section 3.2, a delay time, $t_{delay}$, appropriate for *J-V* measurements can be determined. The rationale is that from the transient device response shown in Figure 3 (a) and (b) and the slow time constants $\tau_{slow}$ extracted, the minimum time for the current transients to become almost congruent can be estimated. We propose that $t_{delay}$ can be either set as a multiple of $\tau_{slow}$ or as the time when $J_{SS}$ in the forward and reverse direction become close to identical. As the $\tau_{slow}$ for the devices under investigation are quite similar and in the order of 0.5 s, we found a minimum delay time of 1 s to be appropriate to carry out J-V measurements.

In Figure 4, *J-V* scans in forward (F, dashed lines) and reverse (R, solid lines) scan directions are compared for $t_{delay}$ of 0.1 ms (colored) and $t_{delay}$ of 1 s (black) for the p-i-n (a) and n-i-p (b) device. *J-V* scans were performed at 50 mV voltage steps and scan conditions are hence equivalent to scan rates of 167 V/s and 0.05 V/s according to the definition in Table S1. Measurements at all *J-V* scan conditions defined in Table S1 were carried out on the p-i-n and n-i-p device and are detailed in the supporting information.

Indeed, there is a substantial discrepancy between the forward and reverse scan for $t_{delay}$ of 0.1 ms while scans performed at a minimum $t_{delay}$ of 1 s, estimated from the transient *MPP* analysis in the previous section, leads to *J-V* curves with small deviation between the scan directions. The *MPP* determined from *J-V* measurements at $t_{delay}$ of 1 s coincide reasonably well with the *MPP* determined from maximum power point tracking data shown in Figure 1. As shown in the SI, the $t_{delay}$ of 1s determined as suitable from the transient analysis shown in Figure 3 was suitably slow for both devices. For the n-i-p device, however, longer $t_{delay}$ of 3 s led to more rather than less discrepancy between the *J-V* curves in different scan directions, indicating that devices may also be altered due to voltage scanning and $t_{delay}$ should be chosen to be sufficiently long but longer $t_{delay}$ may lead to results obscured by device degradation and changes in the device during *J-V* scanning.

For the p-i-n device, the forward measurements for the fast scan exhibit a clear capacitive discharge component between -0.1 V and 0 V as the device was involuntarily kept in open circuit right before the measurement. This decrease in current is comparable to the initial drop in current observed in phase I of the *dynaMPP* measurement shown in Figure 1.

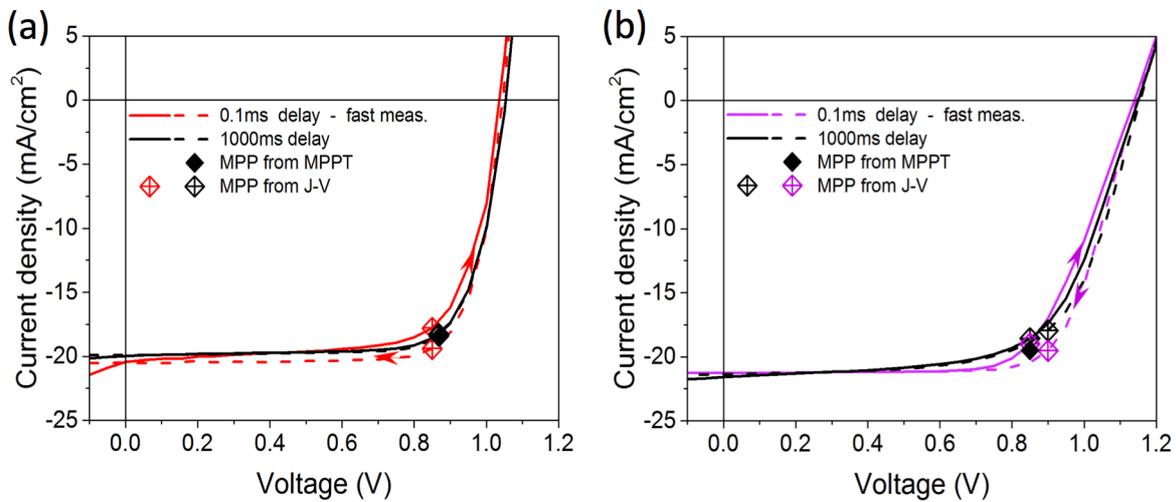

**Figure 4: Current density-voltage, *J-V*, measurements for the p-i-n (a) and n-i-p (b) device performed with short $t_{delay}$ of 0.1 ms (colored) and 1000 ms (black). Measurement directions in forward (F) and reverse (R) directions are marked as solid and dashed lines, respectively, indicated by an arrow. The maximum power points, MPP, determined from the MPP-tracking data shown in Figure 1 and from the J-V measurements shown here are indicated as solid and open/crossed diamonds for all scans and directions.**

Apart from more reliable data on the steady-state performance of metal-halide perovskite based solar cells, the *TrAMPPT* procedure thus enables to directly estimate suitable $t_{delay}$ to perform *J-V* measurements immediately under scan conditions representative of steady-state rather than having to sample many different scan conditions to find those where hysteresis becomes minimal. The latter is time consuming and may not be appropriate for reliable device characterization as reversible transient phenomena cannot be distinguished from device degradation during prolonged measurements.

### 3.4 Comparison of *MPPT* transients with hysteresis indices

To give further evidence, that similar information can be obtained from transient photocurrent response analysis in our proposed *TrAMPPT* procedure and the comparison of *J-V* measurements performed at different *J-V* scan conditions, we compare the differential between the average of the current transients shown in Figure 3 a) and b) with *J-V* scans in different scan directions for the scan conditions defined in Table 1. The results of these measurements can be found in Figure S4 and S6 the SI. The *J-V* discrepancy was quantified by calculating the *HI* as specified in equation (1).

To relate the relative discrepancy between the transient current response in reverse, $J_{trans,R}(t)$, and forward, $J_{trans,F}(t)$, voltage step direction, we calculated the delta difference between the forward and reverse current density transients, $\Delta J(t)$, with respect to the stabilized current density, $J_{SS}$:

$$\frac{\Delta J(t)}{J_{SS}} = \frac{J_{trans,R}(t) - J_{trans,F}(t)}{J_{SS}} \qquad (3)$$

As shown in Figure 5, $\frac{\Delta J(t)}{J_{SS}}$ shows a comparable dependence on time as the *HI* determined from measurements at different delay times. For the p-i-n device, the agreement between the two different sets of measurements is so striking, that we conclude that very similar information can be gaged from the transient analysis around *MPP* and the discrepancy between *J-V* measurements calculated from equation (1) for measurement at different delay times. This is in agreement with the fact that the *J-V* discrepancy is found to be most pronounced around *MPP*.

For the n-i-p device, however, the discrepancy between *J-V* scan directions for longer delay times is larger than suggested by the transient response. As shown in Figure 4 b, *J-V* measurements with minimal discrepancy – hysteresis – were obtained using a $t_{delay}$ of 1 s, derived from analysis of the current density transients. That n-i-p solar cell seems to however exhibit an increased *J-V* discrepancy at $t_{delay}$ = 3 s, as shown in Figure S6, table S3 and indicated by the increased *HI* of Figure 5 (b). This is an indication that voltage cycling during *J-V* measurements changes the device and its performance quite dramatically. At 3 s delay time, a *J-V* scan from 0 V to 1 V with 50 mV increments takes about 1 minute. According to Figure 1, during 1 minute the performance of the n-i-p device noticeably decreases indicating that the charge carrier extraction efficiency may change over time.

This shows that the *TrAMPPT* method is a fast and direct mean to assess the dynamic device response as well as steady-state response and performance over a longer period of time. MPP tracking is also a more reliable indicator of the steady-state device performance as *J-V* cycling itself can induce changes in the device detrimental to device performance. This has previously been debated in poling effects on the performance of metal-halide perovskite based devices.[3,6]

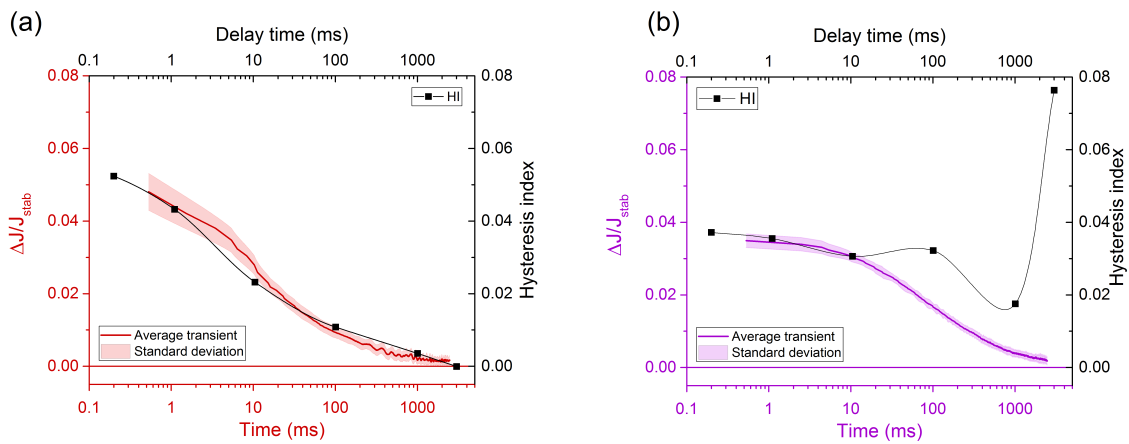

**Figure 5:** Comparison between results of transient device response of the p-i-n (a) and n-i-p (b) devices derived from *TrAMPPT* measurements and *J-V* measurements. For the former, the average difference between the reverse and forward current response, *ΔJ/J$_{stab}$*, calculated according to equation (3) (colored solid trace with deviation) and for the latter hysteresis index, *HI*, calculated according to equation (1) (black squares) is shown to illustrate the similarity and differences between information obtained for different device types.

## 4  Conclusions and Outlook

As a general conclusion of this work, *MPPT* measurements provide more reliable data for the steady-state *MPP* of perovskite devices compared to *J-V* measurements. We propose that *MPPT* data should be provided as a standard when reporting on perovskite solar cell device performance. To some degree, *J-V* measurements should be provided in support of MPPT data to define the device performance metrics of $J_{SC}$, $V_{OC}$ and *FF* and should always be verified to have been measured at scan conditions representative of the steady-state device response.

The *TrAMPPT* procedure proposed here goes a step further by carrying out analysis of the transient device response around *MPP* and from this derive suitable delay times, $t_{delay}$, to perform *J-V* measurements under quasi steady-state conditions. This is of particular importance for solar cells that exhibit pronounced current-transients and change during *J-V* scanning such as metal-halide perovskite solar cells for which *J-V* measurements at a single scan rate do not represent steady-state conditions of the device. We find that the *TrAMPPT* is more time-efficient than having to perform several *J-V* measurements at different $t_{delay}$ to find measurement conditions at which *J-V* discrepancy – hysteresis – become negligible.

We here compare typical p-i-n and n-i-p devices that are representative solar cell architectures showing differences in transient effects with typically less pronounced hysteresis in p-i-n type devices compared to most planar n-i-p architectures, although ionic motion, a key driver for hysteresis was also found in hysteresis-less p-i-n device architectures.[47,48] The example of the p-i-n devices shown here was a device aged for three days, after which it exhibited more pronounced hysteresis than initially, demonstrating that hysteresis may evolve over time due to the creation of ionic defects and imperfections at interfaces. The more detailed analysis presented here shows that hysteresis strongly depends on $t_{delay}$ and devices of different types may exhibit characteristic differences in their transient behavior. The n-i-p device investigated here exhibited similar dynamic response as the p-i-n device but exhibits a more dramatic decrease in photocurrent during *MPPT* measurement indicative of changes in the device causing performance decrease. We will utilize *TrAMPPT* measurements to analyze differences between devices of different architecture types, contact layers, perovskite absorbers and stages in their life-cycle to capture differences, similarities and changes in the *MPP* as well as dynamic response of devices.

We aim to develop the *TrAMPPT* code further and integrate the transient analysis and consecutive *J-V* measurements into the *TrAMPPT* measurement algorithm as illustrated in Scheme 1. This will enable us to "loop" the measurement algorithm and utilize it e.g. for long term stability testing.[29] Apart from the MPP as a function of time, this would enable periodic assessment of changes in the transient device response arising from microscopic changes in the device upon degradation. The slow transient time constant $\tau_{slow}$ can be used to derive $t_{dealy}$ as input parameters for periodic *J-V* measurements, from which a new *MPP* can be defined that then in turn is used as starting *MPP* for the next *MPP* tracking cycles.

**Acknowledgements**

The authors would like to thank Carola Klimm from HZB for technical assistance in SEM measurements and Carola Ferber, Heinz Hagen and Monika Gabernig for technical assistance in the laboratory. A. C. thanks Nga Phung, Philipp Tockhorn and Artiom Bakulin for providing solar cell samples during the course of her MSc thesis work. Laboratory infrastructure in the HySPRINT Innovation Lab has been funded by the Helmholtz Molecular Foundry (HEMF) project. E. U. acknowledges funding from the Swedish Research Council (Project 2015-00163) and Marie Sklodowska Curie Actions Cofund Project INCA (Grant number 600398). E. L. U., K. H., A. C. & J. D. acknowledge funding from the German Ministry of Education and Research (BMBF) for the Young Investigator Group Hybrid Materials Formation and Scaling (HyPerFORME) within the program "NanoMatFutur" (grant no. 03XP0091) and the "SNaPSHoTs" project (grant no. 01IO1806). We further acknowledge the BMBF for funding of the Young Investigator Group Perovskite Tandem Solar Cells within the program "Materialforschung für die Energiewende" (grant no. 03SF0540) and the German Federal Ministry for Economic Affairs and Energy (BMWi) for financial support through the "PersiST" project (grant no. 0324037C). L. K. is a member and acknowledges funding from the graduate school HyPerCell.